\documentstyle[11pt]{article}
\newcommand\beq{\begin{equation}}
\newcommand\eeq{\end{equation}}
\newcommand\bea{\begin{eqnarray}}
\newcommand\eea{\end{eqnarray}}

\begin{document}
\begin{titlepage}
\vfill
\begin{center}
{\Large\bf Strange Form Factors of Octet\\ and Decuplet Baryons$^*$}\\
\vskip 28.mm
{Soon-Tae Hong}
\vskip 0.4cm
{Department of Physics and Basic Science Research Institute,\\
Sogang University, C.P.O. Box 1142, Seoul 100-611, Korea}
\vskip 3.0cm
{\bf ABSTRACT}
\begin{quotation}
The strange form factors of baryon octet are evaluated, in the chiral models
with the general chiral SU(3) group structure, to yield the theoretical
predictions comparable to the recent experimental data of SAMPLE
Collaboration and to study the spin symmetries.  Other model predictions are 
also briefly reviewed to compare with our results and then the strange form 
factors of baryon octet and decuplet are predicted. 
\end{quotation}
\end{center}
\vfill
-----------------------------------------
\par
$^*$Talk given at the 12th Nuclear Physics Summer School and Symposium and the 11th 
International Light-Cone Workshop ``New Directions in QCD," 21-25 June 1999,
Kyungju, Korea
\end{titlepage}

Triggered by the EMC experimental result~\cite{emc} on inelastic muon-proton
scattering, there have been considerable discussions concerning the
strangeness in hadron physics.  Beginning with Kaplan and Nelson's work~
\cite{kaon1} on the charged kaon condensation the theory of condensation in
dense matter has become one of the central issues in nuclear physics and
astrophysics together with the supernova collapse.  The $K^{-}$ condensation
at a few times nuclear matter density was later interpreted~\cite{kaon2} in
terms of cleaning of $\bar{\rm q}$q condensates from the QCD vacuum by a dense
nuclear matter and also was further theoretically investigated~\cite{kaon3} in
chiral phase transition.

Quite recently, the SAMPLE collaboration\cite{sample} reported the
experimental data of the proton strange form factor through parity violating
electron scattering\cite{mck89}.  On the other hand, McKeown\cite{mck} has
shown that the strange form factor of proton should be positive by using the
conjecture that the up-quark effects are generally dominant in the flavor
dependence of the nucleon properties.  This result is contrary to the negative
values of the proton strange form factor which result from most of the model
calculations\cite{jaffe,musolf,koepf,park} except that of Hong and Park
\cite{hp} based on the SU(3) chiral bag model (CBM) and that of Meissner and 
co-workers \cite{meissner}.

Now let us consider the strange form factors of baryons in the chiral models,
such as Skyrmion, MIT and chiral bag models with the general chiral SU(3) group
structure.  In these models, using the electromagnetic (EM) currents obtained
from the model Lagrangian, the magnetic moment operator is given by the sum of
isovector and isoscalar parts, $\hat{\mu}^{i}= \hat{\mu}^{i(3)} 
+\frac{1}{\sqrt3}\hat{\mu}^{i(8)}$ where
$$
\hat{\mu}^{i(a)}
      =
      - {\cal N}D_{ai}^{8}
      - {\cal N}^{\prime} d_{ipq}D_{ap}^{8}\hat{T}_{q}^{R}
      + \frac{N_c}{2\sqrt3}{\cal M} D_{a8}^{8}\hat{J}_{i}
      - {\cal P} D_{ai}^{8} (1-D_{88}^{8})
      + \frac{\sqrt3}{2}{\cal Q}d_{ipq}D_{ap}^{8}D_{8q}^{8}.
$$
Here $D_{ab}^{8}$ is the adjoint representation of SU(3) and
$\hat{J}_{i}=-\hat{T}_{i}^{R}$ $(i=1,2,3)$ and $\hat{T}_{p}^{R}$ $(p=4,5,6,7)$
are the right SU(3) operators along the isospin and strangeness directions,
respectively.  ${\cal M}$, ${\cal N}$ and ${\cal N}^{\prime}$ are the inertia
parameters obtained in the chiral symmetric limit and ${\cal P}$ and
${\cal Q}$ are the inertia parameters coming from the explicit current flavor
symmetry breaking (FSB) effects in the adjoint representation where the chiral
symmetry breaking mass terms cannot contribute to the magnetic moment operator.

Now, in order to take into account the missing symmetry breaking mass effects,
we employ the quantum mechanical perturbative scheme where we use the SU(3)
cranking and treat the symmetry breaking terms perturbatively.  In this
scheme, the Hamiltonian is split up into two pieces, the SU(3) flavor
symmetric and symmetry breaking parts, $H=H_{0}+H_{SB}$ where
\begin{equation}
H_0=M + \frac12 (\frac{1}{{\cal I}_1}-\frac{1}{{\cal I}_2}) \hat{J}^2
        + \frac{1}{2{\cal I}_2} (C_2 (SU(3))
                       - \textstyle\frac34 Y_R^2),~~
H_{SB}=m (1-\hat{D}^{8}_{88}).
\end{equation}
Here $M$ is the static mass of the baryon and ${\cal I}_{1}$ and ${\cal I}_{2}$
are the moments of inertia along the isospin and strangeness directions,
respectively.  $\hat{J}^{2}$ and $C_{2}(SU(3))$ are the Casimir operators in
the SU(2) and SU(3) group, $Y_{R}^{2}$ the right hypercharge operator and $m$
the inertia parameter denoting the symmetry breaking strength.

In the higher dimensional irreducible representation of SU(3) group, the
baryon wave function is described as $|B\rangle = |B\rangle_{8}-C_{\bar{10}}
^{B}|B\rangle_{\bar{10}}-C_{27}^{B}|B\rangle_{27}$ with the representation 
mixing coefficients $C_{\lambda}^{B}$, to yield the implicit FSB contribution 
to the magnetic moment 
\begin{equation}
\delta\mu_{2,B}=-2\sum_{\lambda = \bar{10},27}\frac{
_{8}\langle B|\hat{\mu}^{i}|B\rangle_{\lambda}{}_{\lambda}\langle B|H_{SB}
|B\rangle_{8}}{E_{\lambda}-E_{8}}.\label{delmu2}
\end{equation}
Now the magnetic moments of baryon octet in the FSB case can break up into
three parts
\begin{equation}
\mu_B = \mu_{0,B}({\cal M},{\cal N},{\cal N}^{\prime})
      + \delta \mu_{1,B}({\cal P}, {\cal Q})
      + \delta \mu_{2,B}(m{\cal I}_{2})
\end{equation}
where the first term $\mu_{0,B}$ comes from the chiral symmetric contribution,
$\delta\mu_{1,B}$ is due to the explicit FSB and $\delta\mu_{2,B}$ is obtained
from the implicit FSB in the representation mixing as shown in Eq.
(\ref{delmu2}).

Following the above scheme we can obtain the proton magnetic moment
$$
\mu_{p}=\frac{1}{10}{\cal M}+\frac{4}{15}({\cal N}+\frac{1}{2}
{\cal N}^{\prime})+\frac{8}{45}{\cal P}-\frac{2}{45}{\cal Q}
+m{\cal I}_{2}
(\frac{2}{125}{\cal M}+\frac{8}{1125}({\cal N}-2{\cal N}^{\prime})).
$$
Here one notes that the coefficients are solely given by the SU(3) group
structure of the chiral models and the physical informations such as decay
constants and masses are included in the inertia parameters.

Now one can easily see the spin symmetries as follows.  First, in the adjoint
representation of the SU(3) chiral symmetric limit with ${\cal M}$, ${\cal N}$
and ${\cal N}^{\prime}$, we have the U-spin symmetry, $\mu_{0,p}=
\mu_{0,\Sigma^+}$, $\mu_{0,n} = \mu_{0,\Xi^0}$, $\mu_{0,\Sigma^-}=
\mu_{0, \Xi^-}$ and $\mu_{0,\Lambda}=- \mu_{0, \Sigma^0}.$  Secondly, in the 
implicit representation mixing FSB contributions, we can obtain the V-spin 
symmetry relations, $\delta \mu_{2,p}=\delta \mu_{2,\Xi^-}$, $\delta \mu_{2,n}=
\delta \mu_{2,\Sigma^-}$ and $\delta \mu_{2, \Sigma^+}=\delta \mu_{2, \Xi^0}
 = \textstyle\frac12 \delta \mu_{2,p}$.  Finally, we can see that the s-flavor 
channel possesses the I-spin symmetry, $\mu_{B}^{(s)}=\mu_{\bar{B}}^{(s)}$.

Using the flavor projection operators in the EM currents of the chiral models
we can obtain the strange components of the nucleon magnetic moments, which 
are degenerate in the isomultiplets and respect the above I-spin symmetry
$$
\mu_{N}^{(s)}=-\frac{7}{60}{\cal M}+\frac{1}{45}({\cal N}+\frac{1}{2}
{\cal N}^{\prime})+\frac{1}{45}{\cal P}+\frac{1}{90}{\cal Q}
+m{\cal I}_{2}(\frac{43}{2250}{\cal M}-\frac{38}{3375}({\cal N}
-\frac{13}{19}{\cal N}^{\prime})).
$$

On the other hand, the form factors of the baryons, with internal structure, 
are defined by the matrix elements of the EM currents
$$
\langle p+q|J_{EM}^{\mu} |p\rangle = \bar{u}(p+q)
(\gamma^\mu F_{1B}(q^2)
       +\frac{i}{2m_B} \sigma_{\mu\nu} q^\nu F_{2B}(q^2))u(p)
$$
where $u(p)$ is the spinor of the baryons.  Using the flavor projection
operators in the EM currents as before, in the limit of zero momentum
transfer, one can obtain the strange form factors of the baryons 
\begin{equation}
F^{(s)}_{1B}(0)=S,~~~F^{(s)}_{2B}(0)=-3\mu^{(s)}_{B}-S
\label{f}
\end{equation}
in terms of the strange quantum number of the baryon $S$ and the strange
components of the baryon magnetic moments.  In Table 1, we obtain the chiral 
model predictions that the CBM with $R\approx 0.6$ fm corresponding to 
$\theta (R)=\pi/2$ yield $F_{2p}^{(s)}=0.30$ comparable to the experimental 
data~\cite{sample} $F_{2p}^{(s)}=0.23\pm 0.37\pm 0.15$ (exp) within about 
$30\%$ errors.  Here one notes that the large positive values of the proton 
strange form factors originate from 
$\delta F_{2p}^{(s),1}$ (with ${\cal P}$ and ${\cal Q}$) and $F_{2p}^{(s),2}$
(with $m{\cal I}_{2}$), the explicit and implicit FSB contributions due to
$f_{\pi}\neq f_{K}$, $m_{\pi}\neq m_{K}$ and $m_{u} =m_{d}\neq m_{s}$.

\begin{table}[h]
\caption{The strange form factors of the octet and decuplet baryons.}
\label{table1}
\begin{center}
\begin{tabular}{crrrrrrrrrrr}
$ $ & \multicolumn{1}{c}{$F^{(s),0}_{2N}$}
    & \multicolumn{1}{c}{$\delta F^{(s),1}_{2N}$ }
    & \multicolumn{1}{c}{$\delta F^{(s),2}_{2N}$}
    & \multicolumn{1}{c}{$F^{(s)}_{2N}$}
    & \multicolumn{1}{c}{$F^{(s)}_{2\Lambda}$}
    & \multicolumn{1}{c}{$F^{(s)}_{2\Xi}$}
    & \multicolumn{1}{c}{$F^{(s)}_{2\Sigma}$}
    & \multicolumn{1}{c}{$F^{(s)}_{2\Delta}$}
    & \multicolumn{1}{c}{$F^{(s)}_{2\Sigma^{*}}$}
    & \multicolumn{1}{c}{$F^{(s)}_{2\Xi^{*}}$}
    & \multicolumn{1}{c}{$F^{(s)}_{2\Omega}$}\\
\hline
 CBM  &$-0.19$ &$-0.12$ & 0.61   & 0.30 & 0.49 & 0.25  & $-1.54$  
      &$1.67$ &$0.84$ & 0.56   & 0.83\\
  SM  &$-0.13$ &$-0.09$ & 0.20 &$-0.02$ &0.51 & 0.09   & $-1.74$  
      &$0.04$ &$-0.10$ &$-0.03$ &$0.24$ \\
\hline
\end{tabular}
\end{center}
\end{table}

Now let us briefly review other model predictions.  To the dispersion
theory prediction\cite{jaffe} $F_{2p}^{(s)}=-0.31$, the kaon loop correction is
included\cite{musolf} to yield $F_{2p}^{(s)}=-0.40$ where the SU(3) flavor
symmetric baryon octet, for example $m_{N}=m_{\Lambda}$, is used.  On the
other hand, neglecting the sea-quark fluctuation effects, the nonrelativistic
constituent quark model produces\cite{koepf} $F_{2p}^{(s)}=-0.0324$.  In the
Skyrmion model, Park and collaborators\cite{park} evaluates the proton strange
form factor to yield $F_{2p}^{(s)}=-0.13$, which has the same sign but is much
larger than our Skyrmion prediction due to the fact that they used the
different Skyrmion parameter $e=4.0$ and missed the contribution from the term
proportional to $f_{K}^{2}-f_{\pi}^{2}$ in the inertia parameter 
$m{\cal I}_{2}$.  Very recently Meissner and co-workers\cite{meissner} 
included the kaon loop corrections in the heavy baryon chiral perturbation 
theory to yield $F_{2p}^{(s)}=0.18$, which is positive also.

Similarly to the baryon octet case, one can obtain the magnetic moments of 
$\Delta$ baryons in the s-flavor channel
$$
\mu_{\Delta}^{(s)}=-\frac{7}{48}{\cal M}+\frac{1}{12}({\cal N}-\frac{1}
{2\sqrt{3}}{\cal N}^{\prime})+\frac{2}{21}{\cal P}+\frac{5}{168}{\cal Q}
+m{\cal I}_{2}
(\frac{85}{2016}{\cal M}-\frac{25}{504}({\cal N}-\frac{2}{5\sqrt{3}}
{\cal N}^{\prime})).
$$
Substitution of the above equation into Eq. (\ref{f}) yields the strange form 
factors of $\Delta$ baryons whose numerical values are listed in Table 1, 
together with the predictions for the other octet and decuplet baryons.  

\vskip 1cm
We would like to thank B.Y. Park, D.P. Min, M. Rho and G.E. Brown for helpful 
discussions and constant concerns.


\begin{thebibliography}{99}
\bibliographystyle{unsrt}
\bibitem{emc} Ashman,J., et al., Phys. Lett. {\bf B206}, 364 (1988).
\bibitem{kaon1} Kaplan,D.B., and Nelson,A.E., Phys. Lett. {\bf B175}, 57 
                (1986).
\bibitem{kaon2} Brown,G.E., Kubodera,K., and Rho,M., Phys. Lett. {\bf B192},
                273 (1987).
\bibitem{kaon3} Lee,G.Q., Lee,C.H., and Brown,G.E., Nucl. Phys. {\bf A625},
                372 (1997).
\bibitem{sample} Mueller,B., et al., Phys. Rev. Lett. {\bf 78}, 3824 (1997).
\bibitem{mck89} McKeown,R.D., Phys. Lett. {\bf B219}, 140 (1989);
  Beise,E.J., and McKeown,R.D., Comm. Nucl. Part. Phys. {\bf 20}, 105 (1991).
\bibitem{mck} McKeown,R.D., Los Alamos Preprint {\tt hep-ph/9607340} (1996).
\bibitem{jaffe} Jaffe,R.L., Phys. Lett. {\bf B229}, 275 (1989).
\bibitem{musolf} Musolf,M.J., and Burkardt,M., Z. Phys. {\bf C61}, 433 (1994).
\bibitem{koepf} Koepf,W., Henley,E.M., and Pollock,S.J., Phys. Lett.
                 {\bf B288}, 11 (1992).
\bibitem{park} Park,N.W., Schechter,J., and Weigel,H., Phys. Rev. {\bf D43},
               869 (1991).
\bibitem{hp} Hong,S.T., and Park,B.Y., Nucl. Phys. {\bf A561}, 525 (1993).
\bibitem{meissner} Meissner,Ulf-G., et al., Los Alamos Preprint 
                   {\tt nucl-th/9904076} (1996).
\end{thebibliography}
\end{document}